# On Time Stepping Schemes Considering Switching Behaviors for Power System Electromagnetic Transient Simulation

Sheng Lei
Member, IEEE
Hitachi Energy
San Jose, CA, USA
sheng.lei@hitachienergy.com

*Abstract*—Several difficulties will appear when typical electromagnetic transient simulation, using the implicit trapezoidal method and fixed step sizes, is applied to power systems with switching behaviors. These difficulties are addressed by different aspects of time stepping schemes in the literature. This paper first details the different aspects and reviews corresponding methods. Some misunderstanding in the literature is clarified. Issues that may be encountered by the existing methods are concurrently revealed. Based on the detailed review, the paper then puts forward a novel time stepping scheme which fully addresses the difficulties. The effectiveness of the proposed scheme is demonstrated via numerical case studies.

I. INTRODUCTION

In recent years, more and more power electronic devices are integrated to power grids worldwide, as the interface of FACTSs, HVDCs and renewable energy resources. They are increasingly impacting the transients and dynamics of power systems, and thus receiving growing concerns. These devices introduce numerous frequently switching power electronic valves, resulting in rapidly varying systems, and bringing significant challenges to system studies. Power system studies focus on device-level or system-level behaviors; the detailed transients of individual power electronic valves are not of interest [1]. Therefore, the valves are modeled as idealized switches [1]-[4]. Besides power electronic valves, a fault or an outage can also be understood as a switch.

To study power systems with switches, electromagnetic transient (EMT) simulation is applied. The commonly used EMT simulation tools are based on numerical integration substitution and nodal analysis [5]-[8]. In particular, the implicit trapezoidal method is used to discretize the differential equations, so that the Norton equivalent of circuit elements at a time step is obtained. The nodal equation of the studied system is then built according to Kirchhoff's Current Law (KCL) at each node. A fixed step size is adopted for a simulation run, so that the nodal admittance matrix is constant. This setting avoids the expensive computation required for rebuilding the nodal admittance matrix, and thus greatly enhances the efficiency.

When performing EMT simulation with the implicit trapezoidal method and fixed step sizes on power systems containing switches, the following difficulties appear [1], [4], [9]:
1) How to calculate the actual switching time provided that the switching action may occur between two consecutive time steps?
2) How to determine the combination of all the required switching actions?
3) How to obtain the values of the system at the switching time after all the required switching actions have taken place?
4) How to suppress the notorious numerical oscillation induced by the implicit trapezoidal method after the switching actions [10]-[11]?

These difficulties have been attracting intensive research interest and effort devoted to improving the time stepping schemes for EMT simulation, aspects of which will be detailed in Section II.

Strictly speaking, state variables-based simulation algorithms using variable numerical integrators and step sizes are comparatively more suitable for switched circuits and systems [1], [4], [8]. Nevertheless, their computational efficiency is low and may only be good for small problems [1], [3], [8], [12]-[13]. Moreover, implementing such algorithms requires major changes to the structure of the well-tested existing EMT simulation tools [13]. Using the implicit trapezoidal method and fixed step sizes is still the mainstream in EMT simulation [1], [5]-[6], [8].

The current paper is on time stepping schemes considering switching behaviors. The scope is within offline non-real-time mainstream EMT simulation. Main contributions of the paper are twofold. Firstly, new detailed analysis is provided on the existing methods for different aspects of the time stepping schemes, by applying them to concrete example circuits and carrying out step-by-step calculations. The detailed analysis clarifies some misunderstanding in the literature and reveals issues that may be encountered by the methods. Secondly, a novel time stepping scheme is proposed, which fully addresses the above four difficulties. The effectiveness of the proposed scheme is demonstrated via numerical case studies.

The remainder of this paper is organized as follows. Section II discusses aspects of time stepping schemes considering switching behaviors, with detailed analysis on corresponding existing methods. Section III puts forward the aforementioned novel time stepping scheme. Section IV tests the effectiveness of the proposed scheme via numerical case studies. Finally, Section V concludes the paper and points out some directions for future research.

## II. Aspects of Time Stepping Schemes Considering Switching Behaviors

### A. Switching Time Determination

As mentioned in Section I, existing EMT simulation tools typically adopt the implicit trapezoidal method to discretize the differential equations of the studied power systems, and a fixed user-specified step size $h$ for a whole simulation run. A regular time mesh can be constructed from the start time to the end time of the simulation run, with the fixed step size as the uniform spacing.

Switches are a type of network elements that have discrete operating statuses. A fault resistor can be connected to or disconnected from a power system. A power electronic valve can conduct or stop the current through it. They are examples of switches. A switching action refers to the status change of a switch. Switching actions frequently appear in power systems, usually in the form of fault application/removal, tripping of a device, or turning on/off of a power electronic valve.

It is rather often that a switching action takes place between two consecutive points on the regular time mesh of a simulation run. Nevertheless, digital simulation proceeds in a time-step-by-time-step manner. Switching actions are detected after the calculations for a time step. If no special treatment is taken, the activated switching actions are then applied in the calculations for the next time step. As a result, a switching action may take effect after two time steps since the actual switching time. This time delay in simulation can cause suspicious spikes or non-characteristic harmonics in simulated waveforms [1], [4], [9], [13]-[15].

Calculating the actual switching time is significant for solving the time delay issue, so as to achieve better simulation results. Switching actions can be classified as externally controlled ones and internally controlled ones [3], [12]. In the literature, these two categories may also be called forced commutation and natural commutation respectively [4], [9], [16]. The calculation for the two categories is different, as will be detailed later in this subsection.

Once the switching times of individual switching actions have been calculated, the system switching time $t_{SW}$ can be readily determined. $t_{SW}$ is the switching time of the earliest switching action(s) within the time step that has just been calculated [1], [9], [13]. A sophisticated time stepping scheme will back up the simulation to $t_{SW}$ and then resume it. Note that it is highly probable that the simulation deviates from the regular time mesh after the resumption. An additional step may be taken to back up the simulation to the regular time mesh, which is similar to backing it up to $t_{SW}$ [9], [13]. Nevertheless, this additional step is not mandatory but just optional for offline non-real-time simulation [1], [3].

*1) Externally Controlled Switches and Switching Actions*

The status of some switches and thus the corresponding switching actions are controlled by an external clock [3], [12]. For example, a fault can be applied at a specified time instant; it is disconnected from the power system before that time instant while it is connected after that time instant. As another example, a switch can be opened at a specified time instant; it is on before that time instant while it is off after that time instant.

Typically, the switching times of externally controlled switching actions are known prior to a simulation run. It only remains to check whether or not a given switching time falls within the time step that has just been calculated.

*2) Internally Controlled Switches and Switching Actions*

The status of some switches is controlled by the sign of a certain quantity internal to the studied power system [12], which is called a controlling quantity. Correspondingly, the switching actions are controlled by the zero-crossing of the controlling quantities. For example, when a diode is on, it will remain on if the current through it in the positive direction is greater than 0; it will be turned off once the current drops to zero. Here the current through the diode in the positive direction is the controlling quantity. When the diode is off, it will remain off if the difference between the voltage across it in the positive direction and the forward voltage drop is less than 0; it will be turned on once the difference rises to zero. Here the difference between the voltage across the diode in the positive direction and the forward voltage drop is the controlling quantity.

The switching times of internally controlled switching actions are unknown in advance of a simulation run. They thus have to be calculated during the simulation. In the literature, the switching times of internally controlled switching actions, i.e., the zero-crossing time instants, are usually calculated by assuming linear variation of quantities within the time step that has just been calculated. In other words, linear interpolation is used to approximate the actual waveforms [1], [3]-[6], [9], [13]-[19]. Fig. 1 shows the zero-crossing of a generic controlling quantity $x$. The switching time of the corresponding switching action $t_S$ is calculated as

$$t_S = t - \frac{|x(t)|}{|x(t-h)| + |x(t)|} h \quad (1)$$

where $t$ is the end time of the time step that has just been calculated.

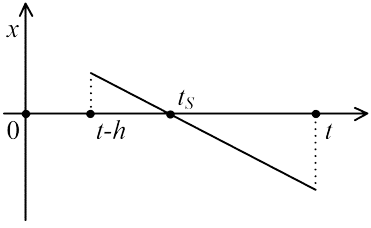

Fig. 1. Zero-crossing of a generic controlling quantity.

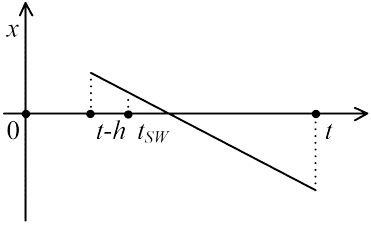

Fig. 2. Linear interpolation of a generic quantity.

Besides linear interpolation, quadratic interpolation can also be used to approximate the actual waveforms [20]. However, quadratic interpolation may not necessarily be more accurate than linear interpolation [16], although it considers more information and performs more calculations.

### B. Interpolation to Switching Time

When $t_{SW}$ has been determined, the quantities of the system at $t_{SW}$ are required to further carry on the simulation. However, the quantities are still unknown at the moment. Linear interpolation is typically adopted to calculate the quantities based on the known values at $t-h$ and $t$ [1], [3]-[6], [9], [13]-[19]. Fig. 2 shows how a generic quantity $x$ is linearly interpolated to $t_{SW}$. According to Fig. 2, the following relation can be constructed

$$x(t_{SW-}) = x(t-h) + \frac{t_{SW} - (t-h)}{h}(x(t) - x(t-h)) \quad (2)$$

Note that $x$ may or may not cross zero at $t_{SW}$. Strictly speaking, the calculated value is that at $t_{SW-}$, namely the time instant immediately before $t_{SW}$, because the calculation is based on the information obtained before the switching action(s) associated with $t_{SW}$.

In addition to linear interpolation, quadratic interpolation can also be used to calculate the quantities at $t_{SW-}$ [20].

### C. Simultaneous Switching

The switching action(s) associated with $t_{SW}$ determined in Section II-A may not be the whole combination of the required switching actions. This is related to simultaneous switching, which refers to the phenomena that one or more switching actions trigger a series of internally controlled switching actions at the same time [1], [3]-[4], [9], [12], [15]-[16], [19]. It should be emphasized that in simultaneous switching, the switching actions do not merely happen to take place at the same time. Instead, they are causally related.

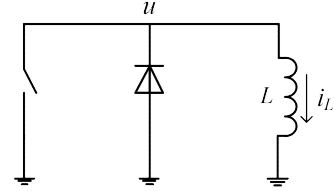

Fig. 3. Example circuit of simultaneous switching.

Simultaneous switching can be intuitively explained via the simple circuit as shown in Fig. 3. It is assumed that the switch is initially closed while the diode is initially off. The diode is ideal: it is an open circuit when it is off, while it is a short circuit when it is on. The initial current going through the inductor is $i_{L0}$. Clearly, the voltage $u$ is initially zero. At the time instant $t_{SW}$, the switch is opened. Theoretically, the diode will be instantaneously turned on to maintain the conductivity of the inductor current. In this example, the opening of the switch triggers the turning on of the diode simultaneously.

A well-designed time stepping scheme should have a mechanism that is able to automatically handle the required switching actions in simultaneous switching, without prior knowledge on the circuit operating mode [3]. To this end, two methods are commonly used in the literature, which will be detailed in the later part of this subsection. Generally, these methods are executed recursively until all the required switching actions are resolved, before the simulation is resumed from $t_{SW}$ [1], [3]-[4], [9], [12], [15]-[16], [19]. Intermediate results are only used to handle the switching actions, but not used in later calculations.

*1) Resolving Simultaneous Switching at Switching Time Using the Implicit Trapezoidal Method*

Simultaneous switching can be resolved at $t_{SW}$ using the implicit trapezoidal method [3], [9], [15]. Applying the implicit trapezoidal method to the inductor at $t_{SW}$

$$i_L(t_{SW+}) = \frac{h}{2L}u(t_{SW+}) + \frac{h}{2L}u(t_{SW}-h) + i_L(t_{SW}-h) \quad (3)$$

Note that the values at $t_{SW+}$ are of interest here, namely the time instant immediately after $t_{SW}$, because they are obtained after the switching action(s). Based on the circuit in Fig. 3, it is reasonable to additionally assume that $u(t_{SW}-h)$ and $i_L(t_{SW}-h)$ are 0 and $i_{L0}$ respectively.

After the opening of the switch, if the diode is still off, the following equation set can be constructed at $t_{SW}$

$$\begin{cases} i_L(t_{SW+}) = \frac{h}{2L}u(t_{SW+}) + \frac{h}{2L}u(t_{SW}-h) + i_L(t_{SW}-h) \\ i_D(t_{SW+}) = 0 \\ i_L(t_{SW+}) - i_D(t_{SW+}) = 0 \end{cases} \quad (4)$$

where $i_D$ is the current through the diode in the positive

direction. Solving (4)

$$u(t_{SW+}) = -\frac{2L}{h}i_{L0} \quad (5)$$

According to (5), there is a positive voltage spike across the diode, which should turn on the diode. Without proceeding in time, the following equation set can be constructed if the diode is on

$$\begin{cases} i_L(t_{SW+}) = \frac{h}{2L}u(t_{SW+}) + \frac{h}{2L}u(t_{SW}-h) + i_L(t_{SW}-h) \\ u(t_{SW+}) = 0 \\ i_L(t_{SW+}) - i_D(t_{SW+}) = 0 \end{cases} \quad (6)$$

Solving (6)

$$i_L(t_{SW+}) = i_D(t_{SW+}) = i_{L0} \quad (7)$$

According to (7), the current through the diode is positive; so the diode will remain on. At this point, no more switching action will take place. Simultaneous switching is resolved automatically by this method.

The following term in (3)

$$\frac{h}{2L}u(t_{SW}-h) + i_L(t_{SW}-h) \quad (8)$$

is the history current of the inductor for $t_{SW}$ [5]-[6]. The history current depends on values prior to $t_{SW}$, which are constant in the calculation for $t_{SW}$ and not impacted by the switching action(s) at $t_{SW}$.

Equations (4) and (6) can be rewritten in the matrix form as (9) and (10) respectively. Here the modified nodal analysis (MNA) is used as the circuit contains ideal elements [21]

$$\begin{pmatrix} \frac{h}{2L} & -1 \\ 0 & 1 \end{pmatrix} \begin{pmatrix} u(t_{SW+}) \\ i_D(t_{SW+}) \end{pmatrix} = \begin{pmatrix} -\left(\frac{h}{2L}u(t_{SW}-h) + i_L(t_{SW}-h)\right) \\ 0 \end{pmatrix} \quad (9)$$

$$\begin{pmatrix} \frac{h}{2L} & -1 \\ 1 & 0 \end{pmatrix} \begin{pmatrix} u(t_{SW+}) \\ i_D(t_{SW+}) \end{pmatrix} = \begin{pmatrix} -\left(\frac{h}{2L}u(t_{SW}-h) + i_L(t_{SW}-h)\right) \\ 0 \end{pmatrix} \quad (10)$$

The right hand sides of (9) and (10) are both a vector consisting of the nodal current injection from the history current for $t_{SW}$. The matrices in the left hand sides of (9) and (10) are both modified nodal admittance matrices at $t_{SW}$ after the opening of the switch. The former is constructed assuming the diode is off; while the latter is constructed assuming the diode is on. The unknown vectors in both (9) and (10) are at $t_{SW+}$. In fact, the two equations can also be constructed by the "Instantaneous Solution" method proposed in [15].

Reference [4] thinks that this method is mathematically questionable. Nevertheless, the derivation in this paper shows that the method is actually mathematically sound. Note that switching actions will not impact history currents, which only depend on values before the switching time. On the other hand, the derivation in this paper reveals a subtle issue with the method: it is not self-starting in that it cannot be directly applied to the first time step of a simulation run. Based on (3), if the method is applied to the first time step, then information prior to the start time of the simulation run is required, which is of course unavailable.

*2) Resolving Simultaneous Switching after Switching Time Using the Backward Euler Method*

Alternatively, simultaneous switching can be resolved at one time step after $t_{SW}$ using the backward Euler method [1], [3], [4], [12]. Ideally, the step size should be small [3], [12]. In the literature, $h/2$ is usually chosen, because the nodal admittance matrix constructed by the backward Euler method with $h/2$ is the same as that constructed by the implicit trapezoidal method with $h$, which saves computation [5]-[6], [11]. Applying the backward Euler method to the inductor at $t_{SW} + h/2$

$$i_L(t_{SW} + \frac{h}{2}) = \frac{h}{2L}u(t_{SW} + \frac{h}{2}) + i_L(t_{SW-}) \quad (11)$$

Note that $i_L(t_{SW-}) = i_{L0}$ is an initial condition. Unlike the method in Section II-C-1, this method needs no additional assumption.

After the opening of the switch, if the diode is still off, the following equation set can be constructed at $t_{SW} + h/2$

$$\begin{cases} i_L(t_{SW} + \frac{h}{2}) = \frac{h}{2L}u(t_{SW} + \frac{h}{2}) + i_L(t_{SW-}) \\ i_D(t_{SW} + \frac{h}{2}) = 0 \\ i_L(t_{SW} + \frac{h}{2}) - i_D(t_{SW} + \frac{h}{2}) = 0 \end{cases} \quad (12)$$

Solving (12)

$$u(t_{SW} + \frac{h}{2}) = -\frac{2L}{h}i_{L0} \quad (13)$$

According to (13), there is a positive voltage spike across the diode, which should turn on the diode. Without proceeding in time, the following equation set can be constructed if the diode is on

$$\begin{cases} i_L(t_{SW} + \frac{h}{2}) = \frac{h}{2L}u(t_{SW} + \frac{h}{2}) + i_L(t_{SW-}) \\ u(t_{SW} + \frac{h}{2}) = 0 \\ i_L(t_{SW} + \frac{h}{2}) - i_D(t_{SW} + \frac{h}{2}) = 0 \end{cases} \quad (14)$$

Solving (14)

$$i_L(t_{SW} + \frac{h}{2}) = i_D(t_{SW} + \frac{h}{2}) = i_{L0} \quad (15)$$

According to (15), the current through the diode is positive; so the diode will remain on. At this point, no more switching action will take place. Therefore, this method is able to resolve simultaneous switching. Moreover, it is self-starting.

*D. Reinitialization*

After the whole combination of the required switching actions has been resolved, the values of the system are to be calculated at $t_{SW+}$, so as to resume the simulation. These values are also utilized to determine possible internally controlled switching actions within the upcoming time step. Reinitialization is the procedure to obtain the numerical approximation to these values.

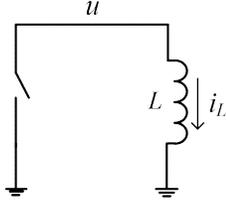

Fig. 4. Example circuit of reinitialization.

On the computational side, it is well-known that the implicit trapezoidal method may induce numerical oscillations after switching actions [5]-[6], [10]-[11]. Another objective of reinitialization is to suppress such numerical oscillations.

In this subsection, existing reinitialization methods are examined with a simple circuit as shown in Fig. 4, to see if the objectives are indeed achieved. It is assumed that the switch is initially closed. The initial current going through the inductor is $i_{L0}$. Clearly, the voltage $u$ is initially zero. At the time instant $t_{SW}$, the switch is opened. Theoretically, $i_L$ should instantaneously change to zero and remain zero afterwards. $u$ should present a negative Dirac impulse; it is negative infinity at $t_{SW}$ while zero elsewhere including $t_{SW-}$ and $t_{SW+}$.

Note that continuity of inductor current does not always hold [12]. To enable Dirac impulses, the continuity has to be relaxed. Simultaneous acceptance of Dirac impulses and insistence on the continuity will lead to self-contradiction, because a Dirac impulse is the derivative of a step function, which is the mathematical characterization of instantaneous change. Moreover, insistence on the continuity will result in violation of KCL. On one hand, the open circuit requires zero current. On the other hand, the inductor current is nonzero along the same loop.

Using the implicit trapezoidal method, the first time step after $t_{SW}$ after the opening of the switch is calculated as

$$0 = i_L(t_{SW} + h) = \frac{h}{2L}u(t_{SW} + h) + \frac{h}{2L}u(t_{SW+}) + i_L(t_{SW+}) \quad (16)$$

$$u(t_{SW} + h) = -\left(u(t_{SW+}) + \frac{2L}{h}i_L(t_{SW+})\right) \quad (17)$$

If a time instant $t$ after the opening of the switch has been calculated, the next time step is calculated as

$$0 = i_L(t + h) = \frac{h}{2L}u(t + h) + \frac{h}{2L}u(t) + i_L(t) \quad (18)$$

$$u(t + h) = -u(t) \quad (19)$$

Note that $i_L(t) = 0$ in (18). According to (17) and (19), numerical oscillations are avoided if and only if

$$\frac{h}{2L}u(t_{SW+}) + i_L(t_{SW+}) = 0 \quad (20)$$

*1) Instantaneous Solution*

The "Instantaneous Solution" method [15], presented in Section II-C-1, calculates the values at $t_{SW+}$. It may also be used for reinitialization. To enable the method, $u(t_{SW}-h)$ and $i_L(t_{SW}-h)$ are additionally assumed to be 0 and $i_{L0}$

respectively. Applying the method

$$\frac{h}{2L}u(t_{SW+}) = -\left(\frac{h}{2L}u(t_{SW} - h) + i_L(t_{SW} - h)\right) \quad (21)$$

Solving (21)

$$u(t_{SW+}) = -\frac{2L}{h}i_{L0} \quad (22)$$

The inductor current is

$$i_L(t_{SW+}) = \frac{h}{2L}u(t_{SW+}) + \frac{h}{2L}u(t_{SW} - h) + i_L(t_{SW} - h) = 0 \quad (23)$$

In the results, $i_L(t_{SW+})$ matches the theoretical value, but $u(t_{SW+})$ does not. Equation (20) is not satisfied, so numerical oscillations will appear. Although the later numerical oscillations can be suppressed via the chatter removal technique [13]-[14] or the critical damping adjustment (CDA) [9], [11], [18], the objectives of reinitialization are not achieved by this method.

*2) Using a Half Time Step of the Backward Euler Method*

A half time step of the backward Euler method may be used for reinitialization [5], [17]. Applying the backward Euler method at $t_{SW} + h/2$

$$0 = i_L(t_{SW} + \frac{h}{2}) = \frac{h}{2L}u(t_{SW} + \frac{h}{2}) + i_L(t_{SW-}) \quad (24)$$

Solving (24)

$$u(t_{SW} + \frac{h}{2}) = -\frac{2L}{h}i_{L0} \quad (25)$$

This method assumes that

$$u(t_{SW+}) = u(t_{SW} + \frac{h}{2}) = -\frac{2L}{h}i_{L0} \quad (26)$$

If the inductor current is approximately considered as linearly varying, then the voltage across it should be a constant. In this situation, (26) makes sense. In addition, the method assumes continuity of inductor current

$$i_L(t_{SW+}) = i_L(t_{SW-}) = i_{L0} \quad (27)$$

The objectives of reinitialization are just partially achieved by this method. Neither $u(t_{SW+})$ nor $i_L(t_{SW+})$ matches the theoretical value. However, (20) is satisfied so there will be no numerical oscillations.

*3) Using Two Half Time Steps of the Backward Euler Method and an Intermediate Linear Extrapolation*

The method proposed in [4] may be used for reinitialization, which uses two half time steps of the backward Euler method and an intermediate linear extrapolation. The values at $t_{SW} + h/2$ are obtained in Section II-D-2.

Linear extrapolation is utilized to calculate $t_{SW} - h/2$

$$i_L(t_{SW} - \frac{h}{2}) = 2i_L(t_{SW-}) - i_L(t_{SW} + \frac{h}{2}) = 2i_{L0} \quad (28)$$

This method assumes continuity of inductor current. Applying a half time step of the backward Euler method from $t_{SW} - h/2$

$$i_L(t_{SW-}) = i_L(t_{SW+}) = \frac{h}{2L}u(t_{SW+}) + i_L(t_{SW} - \frac{h}{2}) \quad (29)$$

Solving (29)

$$u(t_{SW+}) = -\frac{2L}{h}i_{L0} \qquad (30)$$

From the results, this method is essentially the same as that in Section II-D-2. Therefore, the observation on that method also applies to this method.

*4) Using Two Consecutive Half Time Steps of the Backward Euler Method and a Linear Extrapolation*

An alternative method for reinitialization is to use two consecutive half time steps of the backward Euler method and a linear extrapolation [19]. The values at $t_{SW} + h/2$ are obtained in Section II-D-2.

Applying one more half time step of the backward Euler method at $t_{SW} + h$

$$0 = i_L(t_{SW} + h) = \frac{h}{2L}u(t_{SW} + h) + i_L(t_{SW} + \frac{h}{2}) \qquad (31)$$

Solving (31)

$$u(t_{SW} + h) = 0 \qquad (32)$$

By linear extrapolation

$$u(t_{SW+}) = 2u(t_{SW} + \frac{h}{2}) - u(t_{SW} + h) = -\frac{4L}{h}i_{L0} \qquad (33)$$

$$i_L(t_{SW+}) = 2i_L(t_{SW} + \frac{h}{2}) - i_L(t_{SW} + h) = 0 \qquad (34)$$

In the results, $i_L(t_{SW+})$ matches the theoretical value, but $u(t_{SW+})$ does not. Equation (20) is not satisfied, so numerical oscillations will appear. The objectives of reinitialization are not achieved by this method.

*5) Using Forward and Backward Time Steps of the Backward Euler Method*

A classic method in electronic circuit simulation for reinitialization, or for handling inconsistent initial conditions, uses a forward time step and then a backward time step of the backward Euler method [12]. In this paper, the step size is chosen as $h/2$, following the convention in EMT simulation. Again, the values at $t_{SW} + h/2$ are obtained in Section II-D-2.

This method then performs a backward time step with the backward Euler method

$$0 = i_L(t_{SW+}) = -\frac{h}{2L}u(t_{SW+}) + i_L(t_{SW} + \frac{h}{2}) \qquad (35)$$

Note that this method does not insist on continuity of inductor current. Also note that the step size in (35) is negative. Solving (35)

$$u(t_{SW+}) = 0 \qquad (36)$$

In the results, both $u(t_{SW+})$ and $i_L(t_{SW+})$ match the theoretical values. Moreover, (20) is satisfied. This method is the only one examined in this paper that fully achieves the objectives of reinitialization.

III. PROPOSED TIME STEPPING SCHEME

Based on the detailed analysis on aspects of time stepping schemes considering switching behaviors and related methods presented in the previous section, a novel one is proposed here. The mature linear interpolation is still adopted in the proposed time stepping scheme to back up the simulation to the system switching time. The method reviewed in Section II-C-2 is used for resolving simultaneous switching, because it is self-starting. Regarding reinitialization, the method presented in Section II-D-5 is chosen, as it is the only successful one reviewed in this paper.

In particular, the proposed time stepping scheme is detailed as follows. It is assumed that the values of the system at *t-h* have been obtained (*t-h* can be the start time of the simulation run).

1) If *t-h* has reached the end time, terminate the simulation run. Otherwise, solve the system equation set at *t* based on the values of the system at *t-h* using the implicit trapezoidal method.
2) Determine the system switching time $t_{SW}$ within the time step from *t-h* to *t*. All possible switching actions should be checked. For externally controlled switching actions, if the given switching time falls within the time step, one such switching action is considered activated. For internally controlled switching actions, if the corresponding controlling quantity crosses zero within the time step, one such switching action is considered activated; and the switching time is calculated by (1). If there is at least one activated switching action, $t_{SW}$ is equal to the switching time of the earliest activated switching action(s); go to Step 3. If there is no activated switching action, go to Step 1 with the new time setting $t \leftarrow t + h$.
3) Linearly interpolate the values of the system at $t_{SW-}$ using (2).
4) Check and change the status of switches. For externally controlled ones, the status depends on the clock of the simulation. For internally controlled ones, the status is determined by the controlling quantity. If at least one switch should be changed, go to Step 5. If no switch needs to be changed, go to Step 6.
5) Solve the system equation set at $t_{SW} + h/2$ based on the values of the system at $t_{SW-}$ using the backward Euler method. Go to Step 4.
6) Solve the system equation set at $t_{SW}$ based on the values of the system at $t_{SW} + h/2$ using the backward Euler method with a negative step size. The results are the values of the system at $t_{SW+}$.
7) Go to Step 1 with the new time setting $t \leftarrow t_{SW} + h$.

*A. Remarks*

The proposed time stepping scheme can be summarized as: proceeding (Step 1), return (Steps 2 and 3), resolution of simultaneous switching (Steps 4 and 5), reinitialization (Steps 5 and 6), resumption (Step 7). A simulation run is advanced by the proposed scheme with the more accurate implicit trapezoidal method only. The less accurate backward Euler method is merely used in the auxiliary "resolution" and "reinitialization" phases.

Except for the one(s) of which the status should be changed at $t_{SW}$ detected at Step 2, Step 4 does not change other externally controlled switches.

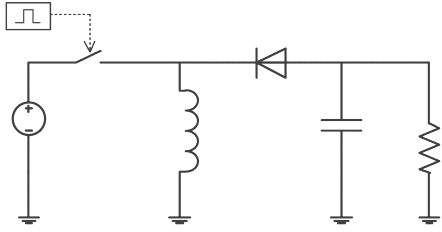

Fig. 5. Buck-boost converter.

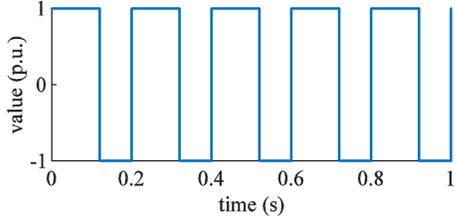

Fig. 6. Square wave.

Steps 4 and 5 are executed iteratively, which compose the method presented in Section II-C-2 for resolving simultaneous switching. The first execution of Step 4 is based on the interpolated values at $t_{SW-}$ obtained from Step 3; while the later executions are based on the solved values at $t_{SW} + h/2$ obtained from Step 5. The last execution of Step 4 ensures that no more switching action should take place, so the workflow moves forward to reinitialization.

The last execution of Step 5 should be accepted, as all the required switching actions have been handled. It constitutes the forward time step of the reinitialization method presented in Section II-D-5. Step 6 completes the method as the backward time step.

### B. Regular Time Mesh

The proposed time stepping scheme itself does not stick to the regular time mesh. That said, if the values of the system on the regular time mesh are of interest, the proposed time stepping scheme will consider the points of the regular time mesh as a special type of externally controlled switching actions. They will trigger the "return" and "resumption" phases of the scheme (Steps 2, 3 and 7) but leave the "resolution" and "reinitialization" phases (Steps 4-6).

## IV. NUMERICAL CASE STUDIES

In this section, the proposed time stepping scheme is tested with numerical case studies on several more complicated power electronic circuits. The proposed scheme has been implemented using MATLAB. Its results are compared to those obtained from Simulink applied to the same test circuits. The comparison demonstrates the effectiveness of the proposed scheme.

The Simulink benchmarking simulations adopt the variable-step solver ode23t with a relative tolerance of $10^{-4}$.

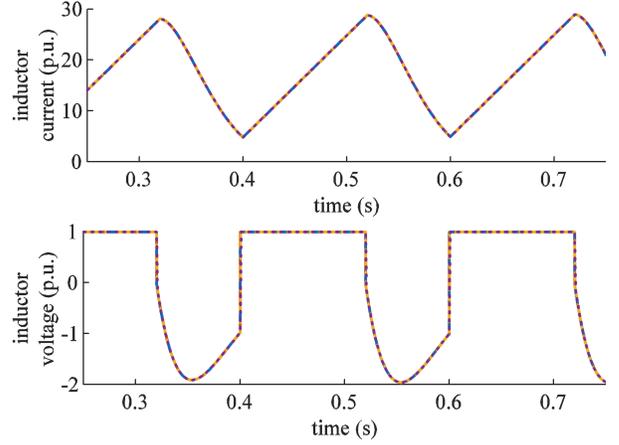

Fig. 7. Results from the continuous operating mode. Solid line: Simulink. Dashed line: proposed, 100 μs. Dash-dotted line: proposed, 500 μs. Dotted line: proposed, 1000 μs.

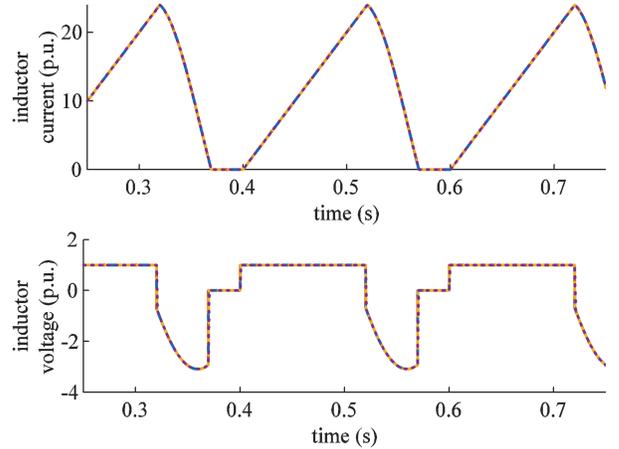

Fig. 8. Results from the discontinuous operating mode. Solid line: Simulink. Dashed line: proposed, 100 μs. Dash-dotted line: proposed, 500 μs. Dotted line: proposed, 1000 μs.

Note that variable-step solvers are considered more suitable for dealing with switching behaviors, as mentioned in Section I. The parameters and variables are presented in per-unit values unless otherwise stated.

### A. Buck-Boost Converter

Fig. 5 shows the circuit of a buck-boost converter. The value of the DC voltage source is 1.0. The inductance is 0.005. The capacitance is 0.2. Two values of the resistance will be considered. 0.1 leads to a continuous operating mode in the inductor current while 0.5 leads to a discontinuous operating mode. The fully controlled switch is driven by a square wave as shown in Fig. 6, the frequency and duty ratio of which are 5 Hz and 0.6 respectively. The switch is on when the level is high while it is off when the level is low.

Simulation results are shown in Fig. 7 for the continuous operating mode and in Fig. 8 for the discontinuous operating mode. The maximum step size in the Simulink simulations is set to 50 μs. The proposed scheme closely matches Simulink in both situations with different step sizes. The curves in each plot basically overlap one another. Note that a 1000 μs step size still produces reasonably high accuracy, although it is unusually large for EMT simulation.

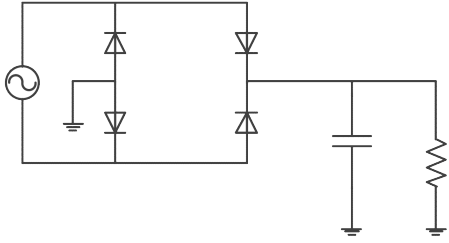

Fig. 9. Single-phase full wave diode rectifier.

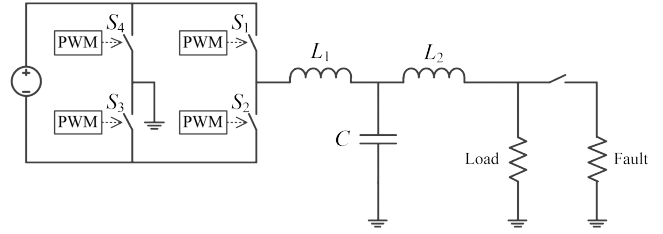

Fig. 11. Single-phase PWM inverter.

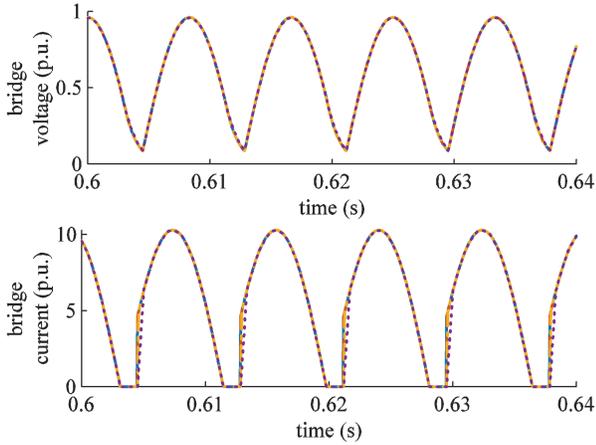

Fig. 10. Results from the single-phase full wave diode rectifier. Solid line: Simulink. Dashed line: proposed, 20 μs. Dash-dotted line: proposed, 100 μs. Dotted line: proposed, 500 μs.

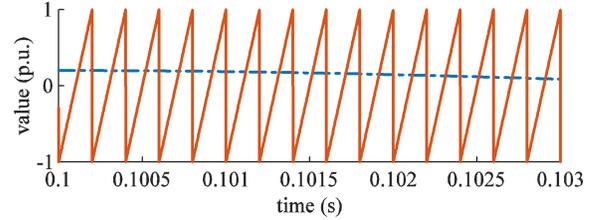

Fig. 12. PWM generation. Solid line: sawtooth wave. Dash-dotted line: targeted sinusoidal wave.

### B. Single-Phase Full Wave Diode Rectifier

The circuit of a single-phase full wave diode rectifier is shown in Fig. 9. The magnitude, initial phase angle and frequency of the AC voltage source are 1.0, π/2 rad and 60 Hz respectively. The capacitance is 0.01. The resistance is 0.1.

Fig. 10 presents the results from Simulink and the proposed scheme. In the Simulink simulation, the maximum step size is set to 10 μs. Again, the proposed scheme accurately agrees with Simulink given different step sizes, including a large one for EMT simulation of 500 μs. The curves in each plot almost overlap one another.

### C. Single-Phase PWM Inverter

This subsection studies a single-phase PWM inverter as shown in Fig. 11 considering a contingency. The value of the DC voltage source is 1.0. In the LCL filter, $L_1$ = 0.0002, $C$ = 0.0005, $L_2$ = 0.0004. The load resistance is 0.06. The PWM signals for the fully controlled switches in the H bridge are generated by comparing a targeted sinusoidal wave and a sawtooth wave. The magnitude, initial phase angle and frequency of the sinusoidal wave are 0.2, π/2 rad and 60 Hz respectively. The sawtooth wave varies between -1 and 1 starting from -1 with a frequency of 5000 Hz. Fig. 12 shows the sinusoidal wave and the sawtooth wave. When the sinusoidal wave is higher than the sawtooth wave, Switches 1 and 3 are on while Switches 2 and 4 are off; otherwise, Switches 1 and 3 are off while Switches 2 and 4 are on. At 0.2 s, a grounded fault is applied at the load terminal with a fault resistance of 0.01. At 0.5 s, the fault is cleared.

The maximum step size adopted by Simulink is set to 1 μs. Results from the proposed scheme with different step sizes are compared to the Simulink results. Fig. 13 presents the inductor currents and capacitor voltage of the LCL filter. Fig. 14 presents the load voltage. The curves from the proposed scheme with different step sizes are rather close to the one from Simulink in each plot, which demonstrates the effectiveness of the proposed scheme. Note that the 50 μs step size still provides reasonable accuracy with the proposed scheme. The step size is a fairly large one compared to the PWM period (200 μs).

### V. CONCLUSION AND FUTURE WORK

Aspects of time stepping schemes for EMT simulation considering switching behaviors and corresponding solution techniques are reviewed in detail in this paper. Through step-by-step analysis on concrete examples, some misunderstanding in the literature is clarified. Concurrently, issues that the existing techniques may encounter are revealed.

A novel time stepping scheme is put forward based on the detailed and critical review. The proposed scheme calculates the actual switching time, simultaneous switching, as well as values of the system at the switching time after the switching actions. The effectiveness of the proposed scheme is tested via numerical case studies.

As a proof of concept, this paper indicates that the proposed scheme is promising and deserves further investigation. Research effort may be directed to quantitative studies on the accuracy and efficiency of the proposed scheme. Applying the proposed scheme to more complicated

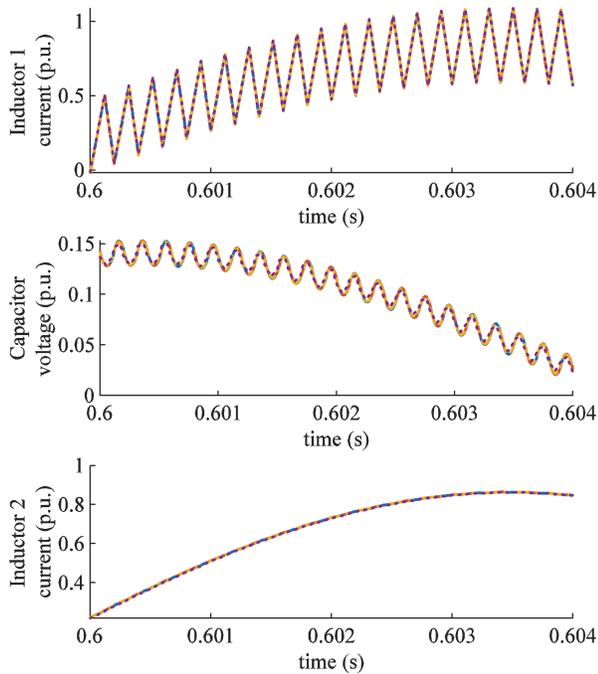

Fig. 13. Inductor currents and capacitor voltage of the LCL filter. Solid line: Simulink. Dashed line: proposed, 10 μs. Dash-dotted line: proposed, 20 μs. Dotted line: proposed, 50 μs.

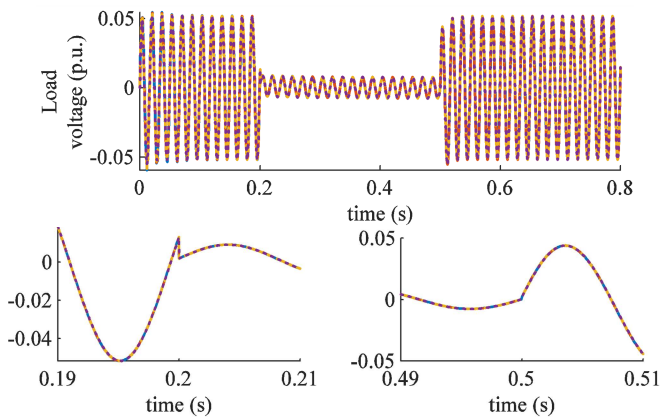

Fig. 14. Load voltage. Solid line: Simulink. Dashed line: proposed, 10 μs. Dash-dotted line: proposed, 20 μs. Dotted line: proposed, 50 μs.

and comprehensive systems with control switching may also be interesting.